\documentclass[12pt]{article}
\newcommand{\bd}{\begin{displaymath}}
\newcommand{\ed}{\end{displaymath}}
\newcommand{\be}{\begin{equation}}
\newcommand{\ee}{\end{equation}}
\newcommand{\bq}{\begin{quote}}
\newcommand{\eq}{\end{quote}}
\newcommand{\ben}{\begin{enumerate}}
\newcommand{\een}{\end{enumerate}}
\newcommand{\bi}{\begin{itemize}}
\newcommand{\ei}{\end{itemize}}
\newcommand{\bdes}{\begin{description}}
\newcommand{\edes}{\end{description}}

\begin{document}
\setlength{\baselineskip}{16pt}
\title{WHY THE LAWS OF PHYSICS ARE JUST SO}
\author{Ulrich Mohrhoff\\
Sri Aurobindo International Centre of Education\\
Pondicherry 605002 India\\
\normalsize\tt ujm@satyam.net.in}
\date{}
\maketitle 
\begin{abstract}
\normalsize\noindent
Does a world that contains chemistry entail the validity of both the standard model of elementary 
particle physics and general relativity, at least as effective theories? This article shows that the 
answer may very well be affirmative. It further suggests that the very existence of stable, spatially 
extended material objects, if not the very existence of the physical world, may require the validity of 
these theories.
\setlength{\baselineskip}{14pt}
\end{abstract}

\section{\large INTRODUCTION}
If at all a fundamental physical theory be derived, it is by teleological arguments---from what it is 
good {\it for\/}. It has been suggested, for instance, that the laws of physics are preconditions 
(conditions of possibility) of empirical science~\cite{vW}. They might, instead, be preconditions of 
observers, of life, or of chemistry. This article presents arguments in support of the view that both 
the standard model of elementary particle physics (SM)~\cite{GGS} and general relativity (GR) are 
preconditions of an ``interesting'' world, defined by Squires~\cite{Squires} as one that contains 
chemistry. It further suggests that the very existence of stable, spatially extended material objects, if not 
the very existence of a physical world, may require the validity of the SM and GR, at least as effective 
theories.

It is well known, but rarely sufficiently appreciated, that matter owes its stability at least in part to the 
indefiniteness of the relative positions between its constituents. Section~2 shows that the proper way of 
dealing with indefinite properties or values 
leads straight to the existence of a unique density operator and the familiar trace rule of quantum 
mechanics (QM). Section~3 explains why the vector space of QM must be complex. Section~4 derives 
the local metric structure of the world, which is described by special relativity. Section~5 
demonstrates that the only possible effects on the motion of a scalar particle are those represented by 
the vector potential~$A$ and the metric tensor~$g$. Section~6 traces the steps leading to quantum 
field theory in general and to quantum electrodynamics (QED) in particular, argues that QED and GR 
are necessary but not sufficient for chemistry, and presents arguments in support of the following: The 
electroweak and strong forces---U(1)$\otimes$SU(2)$\otimes$SU(3)---together with GR constitute 
the simplest theoretical structure consistent with chemistry. Sections~7 and 8 put forward arguments 
suggesting that the very stability of matter, if not the very existence of the physical world, implies both 
the SM and GR.

The final sections address a couple of related issues. Section~8 suggests that the final theory 
envisaged by Weinberg~\cite{Weinberg} and others may be nothing more than the best {\it 
effective\/} theory, and Sec.~9 argues that perhaps we don't need a quantum theory of gravity, 
inasmuch as all that such a theory would allow us to do is investigate the world on scales that do not 
exist.

\section{\large THE ORIGIN OF THE TRACE RULE}
\label{OTR}
An obvious feature of our world is the stability of matter. By this I mean the existence of spatially 
extended material objects that neither explode nor implode the moment they are formed. It is well 
known that matter owes this feature in part to the indefiniteness of the relative positions between its 
constituents; together with the exclusion principle it ``fluffs out'' matter~\cite{Lieb}.

The importance of indefiniteness for the obvious stability of matter can hardly be overstated. This 
makes it an excellent starting point for a derivation of the laws of QM. As is explained 
elsewhere~\cite{Mohrhoff00,Mohrhoff01,RSTAQM}, the proper way of dealing with variables with 
indefinite values is to make counterfactual probability assignments. If an observable is said to have an 
``indefinite value,'' what is meant is that it does not have a value (inasmuch as no value is indicated) 
but that it would have a value if one were indicated, and that positive probabilities are associated with 
at least two possible values.

The most important feature of observables with indefinite values is that their values are {\it 
extrinsic\/}. Since the indefiniteness of an observable implies that it sometimes does and sometimes 
does not have a value, a criterion is called for, and this is the existence of a value-indicating fact---an 
actual event or state of affairs from which the value can in principle be inferred. An observable with 
extrinsic values possesses a value only if, and only to the extent that, a value is indicated by a 
fact~\cite{Mohrhoff00,RSTAQM}. (Since only a fact can indicate something, the words ``by a fact'' 
are of course redundant.) Specifically, {\it no position is a possessed position unless it is an indicated 
position\/}.

What else can we deduce from the existence of observables with indefinite values? Let me begin by 
denouncing a didactically disastrous approach to QM. This starts with the observation that in classical 
physics the state of a system is represented by a point $\cal P$ in some phase space, and that the 
system's possessed properties are represented by the subsets containing~$\cal P$. Next comes the 
question, what are the quantum-mechanical counterparts to $\cal P$ and the subsets containing~$\cal 
P$ {\it qua representations of an actual state of affairs and possessed properties\/}, respectively? 
Once we accept this as a valid question, we are on a wild-goose chase.

If at all we need to proceed from classical physics, the proper way to do so is to point out that every 
classical system is associated with a probability measure, that this is represented by a point~$\cal P$ 
in some phase space, that observable properties are represented by subsets, and that the probability 
of observing a property is~1 if the corresponding subset contains~$\cal P$; otherwise it is~0. Next 
comes the question, what are the quantum-mechanical counterparts to $\cal P$ and the subsets 
containing~$\cal P$ {\it qua representations of a probability measure and observable properties\/}, 
respectively? Once we have the answer to this we are ready for the next question: Is it possible to 
reinterpret the quantum-mechanical counterpart to $\cal P$ as representing an actual state of affairs 
connoting a set of possessed properties? Because the classical probability measure assigns trivial 
probabilities (either 0 or~1), it is possible to think of it as an actual state of affairs. Because quantal 
probability measures generally assign nontrivial probabilities (neither 0 nor~1), it is not possible to 
similarly reinterpret the quantum counterpart to $\cal P$.

To find the quantum counterpart to $\cal P$, all we have to do is make room for nontrivial 
probabilities, and the obvious way to do this is to replace the subsets of a phase space by the 
subspaces of a vector space, or by the corresponding projectors. An atomic probability measure will 
then be a 1-dimensional subspace $U$, probability~1 will be assigned to properties that contain~$U$, 
probability~0 to properties that are orthogonal to~$U$, and the remaining properties will be 
associated with nontrivial probabilities. The possible values of an observable will evidently correspond 
to an orthogonal resolution of the identity, and compatible attributions of properties will be 
represented by projectors that are sums of projectors from an orthogonal resolution of the 
identity---that is, by commuting projectors. Finally, the probability assigned to the sum of two 
orthogonal projectors will be the sum of the probabilities assigned to the individual projectors. This is 
nothing but the classical sum rule for the probability of ``either $A$ or $B$,'' which holds if $A$ and 
$B$ are members of a set of mutually exclusive events, provided that one of them happens. Since 
probabilities are assigned on the proviso that a value be indicated, this condition is always satisfied.

All of the above follows directly from the obvious way to make room for nontrivial 
probabilities---namely, the substitution of subspaces for subsets and 1-dimensional subspaces for 
points. It is sufficient to prove Gleason's theorem~\cite{Gleason,Pitowsky} for real and complex 
vector spaces with more than two dimensions~\cite{Peres95}, according to which probability 
assignments based on earlier value-indicating facts are governed by the familiar trace rule and a 
unique density operator. The theorem has recently been proved for two dimensions as 
well~\cite{Fuchs,Busch,Fivel}. The question as to why QM needs a {\it complex\/} vector space will 
be answered in the following section.

\section{\large WHY COMPLEX NUMBERS?}
\label{WCN}
Consider a series of measurements. (A measurement is anything that results in an actual event or 
state of affairs from which the possession of a property by a physical system or a value by a physical 
variable can in principle be inferred.) If we assign probabilities to the possible results of the final 
measurement, we have to do so in conformity with the probability algorithm derived in the previous 
section. If the intermediate measurements are not performed but the histories that lead to a possible 
result of the final measurement are defined in terms of the possible results of the intermediate 
measurements, the probability of any possible final result will, according to that algorithm, be given by 
the (absolute) square of a sum over all the histories that lead to this result. Each history contributes 
an amplitude, which may be real or complex.

Next consider the limit of a series of unperformed position measurements on a scalar particle, in which 
the histories become continuous trajectories. Suppose $s$ parametrizes such a trajectory, and $ds_1$ 
and $ds_2$ label adjoining infinitesimal segments. Our probability algorithm requires us to multiply 
the amplitudes associated with successive segments of a history, so that
\be
A(ds_1+ds_2)=A(ds_1)A(ds_2).
\ee
Hence the amplitude for propagation along an infinitesimal path segment can be written as 
$A(ds)=\exp(z\,ds)$, and the amplitude for propagation along a path $\cal C$ can be written as 
$A(s[{\cal C}])=\exp(zs[{\cal C}])$. If $z$ had a real part, the probability of finding the particle 
anywhere would not be conserved; it would either increase or decrease exponentially with time. It 
follows that $A(s[{\cal C}])$ must be a phase factor $\exp(ibs[\cal C])$.

\section{\large THE ROAD TO SPECIAL RELATIVITY}

Consider a scalar particle and a particular continuous path. As the particle travels along this path (in 
our imagination if nowhere else), $s$~increases, and $\exp(ibs)$ rotates in the complex plane. Let us 
say that every time it completes a cycle, the particle ``ticks.'' If the particle is free, it singles out a 
class of uniform time parameters---those for which the number of ticks per second is constant. 
Different particles may tick at different rates, which are related to the standard rate of one tick per 
second by the species-specific constant factor~$b$.

So much for the origin and meaning of mass and proper time. Our next task is to determine the 
physical roots of the spatial part of the metric. An analysis of two-slit interference experiments (in the 
framework of standard QM)~\cite{WATQM,AK} has shown that physical space cannot be a 
self-existent and intrinsically differentiated expanse. Space is the totality of positions that are 
possessed by material objects, and since these are relatively defined, there are no absolute positions. 
By the same token, there is no absolute rest. As a consequence, the proper-time interval $ds$ and 
inertial coordinates are related via
\be
ds^2=dt^2+K(dx^2+dy^2+dz^2),
\ee
where $K$ is a universal constant, which may be positive, zero, or negative~\cite{SU}.

Here are some of the reasons why $K>0$ can be excluded. (i)~Ubiquitous causal loops: Reversing an 
object's motion in time is as easy as changing its direction of motion in space. (ii)~The nonexistence 
of an invariant speed---a speed that is independent of the inertial frame in which it is 
measured---rules out massless particles, long-range forces, and the possibility of resting the spatial 
part of the metric on the cyclic behavior of particles (the rates at which they tick). The last point is 
decisive, in as much as all there is to fix the spatial part of the metric is the rates at which free or 
freely falling particles tick. Since space is not a self-existent and intrinsically differentiated expanse, 
the metric has to be defined---as well as brought into existence---by the behavior of the world's 
material constituents. And the only thing that a scalar particle does is tick as time passes. This cyclic 
behavior realizes (makes real) the temporal part of the metric. Hence if there is to be a spacetime 
metric, its spatial part must be determined by its temporal part.

If $K$ is not positive, causal loops are ruled out by the existence of an invariant speed. For $K=0$ this 
is infinite, and for negative $K$ it is $c=\sqrt{1/|K|}$. The problem with the nonrelativistic case 
$K=0$ is that the rates at which free particles tick still cannot fix the spatial part of the metric. They 
just define a universal inertial time scale via $ds=dt$. Nor are light signals available for converting 
time units into space units. Nor do we get interference from free particles since $\exp(ibs[{\cal C}])$ 
is the same for all paths with identical endpoints---and without interference QM is inconsistent (with 
the existence of a macroworld, which it presupposes~\cite{Mohrhoff00,RSTAQM,WATQM}). It is 
possible to obtain interference via an action that depends on the frame, but if all there is to fix the 
spatial part of the metric---for every inertial frame---is the cyclic behavior of free or freely falling 
particles, this ought to be invariant. And for this we need a finite invariant speed (negative $K$).

\section{\large THE CLASSICAL FORCES}
\label{CF}
The rates at which particles ``tick'' not only found the metric properties of the world but also make it 
possible to influence the behavior of matter by influencing particle propagators. The only way of 
influencing the probability of finding at one spacetime location a scalar particle last ``seen'' at another 
location, is to modify the rate at which it ticks as it travels along each path connecting the two 
locations. The number of ticks associated with an infinitesimal path segment defines a species-specific 
Finsler geometry 
$dS(dt,d{\bf r},t,{\bf r})$~\cite{Franz,Finsler}. As the following will show, there are just two ways of 
influencing the Finsler geometry that goes with a scalar particle.

As mentioned, the stability of matter rests on both the indefiniteness of relative positions and the 
exclusion principle~\cite{Lieb}. For the exclusion principle to hold, the ultimate constituents of matter 
must be indistinguishable members of one or several species of fermions. The necessary 
indistinguishability entails that all free particles belonging to the same species of fermions tick at the 
same rate. This guarantees the possibility of a global system of spacetime units~\cite{MW}: While 
there may be no global inertial frame, there will be local ones, and they will mesh with each other as 
described by a Riemannian spacetime geometry. Accordingly there is a species-specific way of 
influencing the Finsler geometry associated with a scalar particle, which bends geodesics relative to 
local inertial frames, and there is a species-independent way, which bends the geodesics of the 
Riemannian spacetime geometry. In natural units:
\be
dS=m\sqrt{g_{\mu\nu} dx^\mu dx^\nu}+qA_\nu dx^\nu,
\ee
where $A$ and $g$ represent the two possible kinds of effects on the motion of scalar particles.

If the sources of these two fields have no definite positions, the fields themselves cannot have definite 
values. We take this into account by summing over histories of $A$ and~$g$, and for consistency with 
the existence of a macroworld (presupposed by QM~\cite{Mohrhoff00,RSTAQM,WATQM}) we make 
sure that a unique history is obtained in the classical limit. Obvious and well-known constraints then 
uniquely determine the terms that need to be added to~$dS$ (except for a possible cosmological 
term)~\cite{LL2}.

We have reached an inconsistency, which no physical theory proposed so far has been able to 
eliminate. Since we sum over histories of~$g$, none of the continuous paths contributing to a particle 
propagator has a definite spacetime length. And since the spacetime lengths of these paths constitute 
the physical basis of the metric properties of the world, definite distances or durations do not exist. 
But the spacetime points on which $g$ as a field depends can only be defined by the spacetime 
distances that exist between them. If definite spacetime distances do not exist, the manifold over 
which field theories are constructed does not exist either. (Renormalization is a way of glossing over 
this inconsistency.)

\section{\large THE STANDARD MODEL}

Huygens' principle tells us that a sum over spacetime trajectories can be replaced by a wave equation. 
A relativistic wave equation has ``negative energy'' solutions corresponding to particles for which 
proper time decreases as inertial time increases~\cite{CMR}, and it conserves charge rather than 
probability. Particle numbers are therefore variables, and since the quantities that dynamically 
influence their values are fuzzy, so are they.

To accommodate fuzzy particle numbers we sum over the histories of a field the Lagrangian of which 
yields the wave equation in the classical limit. This turns the field modes into harmonic oscillators 
whose quanta represent individual particles with definite energies and momenta. Expanding the 
interaction part of $\exp(iS)$ yields a sum over histories in which free particles are created and/or 
annihilated, and by using the appropriate wave equation for spin-$1/2$ particles we arrive at the 
Feynman rules for QED. (The wave equation for a free particle is obtained by imposing 
on the wave function a specific relation 
between the energy $E$ and the absolute value $p$ of the momentum. This too is a 
consequence of the fact that the spatial part of the metric is determined by its temporal part---the rates 
at which free particles tick. For spin-$1/2$ particles the relativistic relation between $E$ and $p$ leads to 
the Dirac equation. A half-integral spin is required by the stability of matter since the exclusion principle 
only holds for fermions.)

Gravity and electromagnetism are clearly necessary for a world that contains something as interesting 
as chemistry. Without the electromagnetic force there would be no Periodic Table, and without gravity 
there would be no stars to synthesize the Table's ingredients beyond helium. But the two classical 
forces are not sufficient. Nucleosynthesis calls for another force. The requirement of renormalizability 
pretty much narrows this down to a non-Abelian gauge field. {\it A posteriori\/} we know that three 
colors suffice. Why not less? One reason is that two would not lead to confinement. There would be no 
nucleons for making a variety of nuclei and no residual force that is both attractive and 
short-range---capable of binding nucleons into nuclei and incapable of spoiling the delicate 
electromagnetic equilibria on which the Periodic Table depends. Thus three colors are needed as well as 
sufficient, which fixes the Lagrangian for quantum chromodynamics~\cite{MP}, give or take a few 
adjustable parameters.

Yet another force is needed, for several reasons. If stars did not explode, the elements beyond helium 
would remain confined to the interiors of stars, where they are created. The force responsible for 
stellar explosions---the weak force---has the simplest non-Abelian gauge group. It also plays an 
important role in nucleosynthesis and critically controls the primordial fusion of hydrogen into helium. 
Since SU(2) does not lead to confinement, and since the stability of matter forbids the weak force to 
cause the decay of atomic electrons through interactions with nuclear quarks, a different mechanism 
for reducing its range is needed---the Higgs mechanism~\cite{AL}. Moreover, the electron's 
flavor-doublet partner must be neutral if the weak force is to trigger stellar explosions~\cite{BT}.

Thus once again one is left with little choice. The non-conservation of parity remains something of a 
mystery, as does the triplication of flavor pairs, but there are indications that both are needed to 
create the required excess of matter over antimatter~\cite{GGS,SS}. (The total annihilation of both 
matter and antimatter would spell the total annihilation of all spatiotemporal relations and thus the 
total annihilation of space and time.)

While these arguments are certainly not rigorous, they nevertheless make it plausible that the SM and 
GR together constitute the simplest theoretical structure consistent with a world that contains 
chemistry. It seems that we do live in the simplest possible interesting world~\cite{Squires}. In the 
following two sections I shall submit stronger claims, namely that the stability of matter may call for a 
world in which the SM is at least an effective theory (Sec.~7), and that such a world may be the only 
possible physical world (Sec.~8).

\section{\large THE STABILITY OF MATTER AND THE STANDARD MODEL}

The ``anthropic'' principle~\cite{BT,Hogan} is usually invoked to account for (i)~the approximate 
values of some of the SM's 19$+$ adjustable parameters and (ii)~certain rather special cosmological 
initial conditions. What the previous sections have shown, or strongly suggest, is that the entire SM 
(including GR and the general theoretical framework of quantum field theory) has teleological 
underpinnings. It not only describes almost all known physical phenomena (exceptions are the 
recently observed neutrino masses, if confirmed, and the as yet unknown nature of dark 
matter~\cite{Sadoulet}) but it also appears to contain nothing but the simplest set of physical laws 
permitting the existence of something as interesting as chemistry.

Laws that are at least effectively identical with the SM may follow from a weaker requirement. The 
stability of matter (as defined at the beginning of Sec.~\ref{OTR}) rests on the indefiniteness of 
relative positions. The search for the right formalism for dealing with indefinite variables takes us 
straight to~QM. But QM would be inconsistent without macroscopic objects, inasmuch as no position is 
a possessed position unless it is an indicated position (Sec.~\ref{OTR}), and only a macroscopic 
object can indicate something. (The reason why only macroscopic objects can be indicators is that 
only the positions of macroscopic objects can be consistently thought of as forming a self-existent 
system of causally connected properties that are effectively detached from the facts by which they are 
indicated~\cite{Mohrhoff00,RSTAQM,WATQM}.) The stability of matter thus entails the existence of 
macroscopic objects in general and of macroscopic detectors in particular. But it may well be that 
macroscopic detectors can exist only in worlds that contain a sufficient variety of chemical elements. If 
this turned out to be the case, the stability of matter would require chemistry. Hence it would entail 
physical laws that are (at least effectively) identical with the SM (including GR). In short:
\bd
\hbox{Stable Matter}\longrightarrow\hbox{QM}\longrightarrow\hbox{Detectors}
\stackrel{\textstyle?}{\longrightarrow}\hbox{Chemistry}\longrightarrow\hbox{SM+GR}
\ed

\section{\large THOUGHTS ON A FINAL THEORY}

In the context of the SM only certain quite unlikely values of the adjustable parameters and the 
cosmological initial conditions are consistent with the phenomenology encapsulated in the Periodic 
Table. To some this suggest design, to others it suggests the existence of an ensemble of universes 
across which parameters and initial conditions take random values, while to the majority of theoretical 
physicists it suggests the need for a final theory which allows most of the actual parameter values to 
be derived from first principles, and which either stipulates the initial conditions~\cite{HaHa} or 
makes the universe largely insensitive to them~\cite{GuStein}.

I can see another possibility. Supposing that chemistry is indeed required for the existence of 
detectors (and thus for the stability of matter), a world governed by laws that are not (effectively) 
identical with the SM will be void of detectors, as will be a world with the ``wrong'' parameter 
values and/or cosmological initial conditions. But a world without detectors is a world without 
positions, and a world without positions is a spaceless world. And since a spaceless {\it physical\/} 
world is a contradiction in terms, we can conclude that {\it any\/} physical world will (at least 
effectively) be governed by the SM and GR. For the same reason we cannot conceive of universes with 
the ``right'' parameters and initial conditions as members of an ensemble across which parameters 
and/or initial conditions vary. Universes with the wrong parameters or initial conditions are logically 
inconsistent and therefore nonexistent. Nor can we infer design if only worlds with the right 
parameters and initial conditions are logically consistent.

What about the possibility of a final physical theory that leaves no room for ``anthropic'' adjustments? 
If the cognitive apparatus of the human mind is up to the task of discovering the true structure of the 
world, we can expect that this will eventually be discovered. And if this structure---or a part of 
it---admits of a precise mathematical description, it may well turn out to be faithfully represented by 
such a theory. But the human mind may not be up to that task, and the true structure of the world 
may not be mathematical. In either case {\it all\/} physical theories would be effective approximations 
to the true structure of the world, which would remain forever beyond our ken. There could still be a 
final theory, but it would simply be the best effective theory, and it could contain as reminders of its 
effective nature such inconsistencies as the one pointed out at the end of Sec.~\ref{CF}.

Elsewhere~\cite{RSTAQM,WATQM} I argued that the world is constructed from the top down, by a 
finite spatial differentiation that stops short of realizing the multiplicity of a manifold, rather than built 
from the bottom up, on an intrinsically and maximally differentiated manifold, out of locally 
instantiated properties. There are psychological as well as neurobiological 
reasons~\cite{BCCP,QMCCP} to believe that the physically inadequate bottom-up approach is intrinsic 
to the human mind. If so then it appears that we cannot hope for more than an optimal effective 
theory.

\section{\large A QUANTUM THEORY OF GRAVITY?}

The search for a final theory is in large measure a search for a quantum theory of gravity. Yet there 
are reasons to believe that the idea of such a theory is self-contradictory, as the following will show.

As a rule, the indefiniteness of a variable cannot evince itself without something less indefinite. The 
indefiniteness of the position of a material object can evince itself (through statistical fluctuations) 
only to the extent that detectors with sharper positions exist. The residual fuzziness of the positions of 
macroscopic objects cannot evince itself, which is why these positions are effectively 
intrinsic~\cite{Mohrhoff00,RSTAQM,WATQM}.

Again, the indefiniteness of the electromagnetic field can evince itself to the extent that distances are 
well-defined. The fuzziness of~$A$ induces an indefiniteness in the species-specific lengths 
(Sec.~\ref{CF}) of the histories (paths) associated with electrically charged particles, and so does the 
fuzziness of~$g$. On observationally accessible scales the indefiniteness induced by $A$ far exceeds 
that induced by~$g$. Down to scales on which $g$ becomes as fuzzy as $A$, distances are 
well-defined. This is why the fuzziness of~$A$ has observables consequences such as the Lamb shift. 
(This effect exists not only because a 2S electron is closer to the proton on average than is a 2P 
electron but also because on atomic scales ``closer'' is still extremely well defined.)

On the other hand, there is nothing less fuzzy than the metric. This suggests to me that there are no 
physical effects that are due to the indefiniteness of the metric, just as there are no physical effects 
that are due to the indefiniteness of the positions of macroscopic objects. If so, a quantum theory of 
gravity would allow us to study the physical consequences of a fuzziness that lacks physical 
consequences. Such a theory would make it possible to investigate the physics on scales that do not 
exist, for on scales on which the fuzziness of $g$ becomes significant, the concept of ``scale'' loses its 
meaning.

\end{document}